# Inverse Occam's razor

Igor Mazin, George Mason University, Fairfax, VA, USA
imazin2@gmu.edu

**Scientists have long preferred the simplest possible explanation of their data. More recently, a worrying trend to favor complex interpretations has taken hold because they are perceived as more impactful.**

One of the fundamental principles in science is the law of parsimony. It is usually (but probably incorrectly) attributed to William of Ockham, a 14[th] century English and therefore colloquially referred to as Occam's razor. Of many equivalent formulations of said principle, I personally favor the following: *Of two competing theories, the simpler explanation of an entity is to be preferred*[1].

For centuries, starting from Galileo, this credo has been a cornerstone of the scientific method. In the last decades, it seems, it has been supplanted by the opposite maxim: *of two competing interpretations, the more exotic one is to be preferred. In a way, it is but human nature, to chase after something more intriguing, less quotidian – but that factor has been in place for ages. What is at work here and now is a more pragmatic assumption - that a more exciting interpretation can get your paper published in a high-profile journal.*

Let us consider some examples (I intentionally give no explicit citations, for their name is Legion, and I see no reason to single out one or two affronts). Suppose you have measured a transport effect (such as anomalous Hall conductivity or linear magnetoresistance), which can, in principle, be attributed to Dirac bands – or to other bands, which are often present in the materials concurrently with the Dirac electrons, and carry a large part of the electron transport. You can bet 10 to 1 that the authors will push for the former explanation.

An excellent example of how destructive this tendency may be to science is the infamous putative *p*-wave superconductor, $Sr_2RuO_4$. After initial NMR experiments seemingly indicated possible triplet pairing (a phenomenon much more exotic than conventional superconductivity, or even the *d*-wave superconductivity in high-$T_c$ cuprates), the physics community was mesmerized by this possibility. As evidence was being accumulated against this enthralling interpretation, it was simply neglected and swept under the rug. This went on for two decades, until in 2020 it was directly and conclusively proven that the original NMR experiments were flawed (to their credit, the flaw was subtle and unexpected), and correct measurements exclude a triplet state with a high degree of confidence.

I have a neat little collection of similar examples. Sometimes the exotic explanation would reign for a year, sometimes for a few years, sometimes for decades. Sometimes it is still on the table, despite several papers pointing toward a prosaic explanation. As a rule ($Sr_2RuO_4$ is a notable exception), while the original paper would be published in a high-impact journal such as *Nature* or *Science*, the one debunking it would go to a more quotidian venue such as *Phys. Rev.* And, of course, it would never accumulate as many citations.

For example, quite some time ago an experimental paper was published in *Nature* declaring a highly exotic case of coexistence of unconventional superconductivity and magnetism. For the next 5 years or so, it was cited at a rate of 40-50 times a year. Then, a rather detailed theoretical paper appeared, with quantitative calculations, indicating that the suggested interpretation did not seem to be internally consistent. This paper was published in Phys. Rev. B and cited 10 times over the following 5 years (and ~20 more times since) – despite the fact that less than a year later the authors of the original paper revisited their experiment and found, to their credit, that the original observation was due to extrinsic effects and the conclusion was invalid. This paper was also published in Phys. Rev. B and cited less than 40 times in its lifetime. Amazingly, the original publication, already known to be incorrect, was, in the same time period, cited 200 times!

This phenomenon of inverse Occam's razor is intimately related to another relatively novel tendency: an experimental paper that reports observations without an interpretation has little chance to fly high. I recall a special issue of *Nature* collecting the most influential *Nature* papers of the 20$^{th}$ century. One of those was Pyotr Kapitza's letter[2] reporting his discovery of superfluidity. It was one page long, and did not offer any attempt at an explanation. It was simply stating that immeasurably small viscosity was observed in helium below the λ-point. It was a Nobel Prize paper[3].

Those were, of course, different times, as exemplified by another brilliant quote, attributed either to Irène Joliot-Curie or to her husband (probably incorrectly in either case) is, *The farther an experiment is from theory, the closer it is to the Nobel Prize.* Nowadays, if your experimental paper does not provide any theoretical speculation (sometimes "supported" by first principles calculations that in reality do not support anything of the sort, but only show that you can run a canned code, and sometimes by analytical model that have hardly anything to do with the actual material, but generate a plethora of buzz words, it is doomed to fail the scrutiny of *Nature* or *Science* reviewers and editors.

This interestingly coexists with another, much older trend that stems from the ancient times when experiment was seen as the king crowned by Galileo and theory as a deposed pretender, a mere priest ordained by Aristotle.

Let me take an excursion down the memory lane to illustrate this point: I remember how in the late 70s I and a few other young theorists at the Lebedev Institute in Moscow were assigned a new office to share, and found, to our dismay, that it lacked a blackboard. We immediately placed a job order, to remove what we saw as useless pipes sticking out of the wall, and hang a new blackboard. The gentleman who promptly arrived to access the job immediately refused, saying that these pipes supply oxygen and hydrogen and cannot be removed. "But we are theorists", we cried, "we do not need hydrogen!" – "Not my business", he retorted. "Today there are theorists in this office, tomorrow physicists will move in."

This attitude has plagued us theorists at Science and Nature for long time, and still does, albeit to a lesser extent, even though I personally think it outdated. In these journals, it is definitely easier to publish an experiment without theory than a theory without experiment. As counterintuitive as I think it is, these journals are also less likely to accept a theory that explains already published, but not yet understood — or incorrectly interpreted — experiments. Admittedly, it is often harder to referee a pure theory paper and harder to anticipate possible impact, but hard choices should not necessarily be avoided.

We have all seen idealistic treatises on perceived attitude problems in physics (and in science in general) — from overreliance on the much vilified citation indices, to the perils of anonymous refereeing. The authors usually end their diatribes with unrealistic pleas to avoid these wicked practices voluntarily. They urge their readers not to pay attention to citations when hiring new faculty, not to count high-impact journals in the applicant's CV, to sign your peer reviews… Well, as an old Yiddish saying states: *it's better to be rich and healthy than to be poor and sick*. That is to say, it would be great to live a utopic dream, but better not count on it.

I want to offer a simple and realistic path to fighting these predicaments, which, I am confident, is very practical. First, do not punish authors for not having a theoretical model for their observations. The usual argument that "the paper is not of sufficiently general interest" is often a veiled way of saying that the authors have not adorned their experiment with sufficiently fancy theoretical verbiage. Forget about that. If the result does not immediately lend itself to an interpretation it should be considered an asset, not a shortcoming. The scientific community gets a new challenge to solve and the authors may even be "closer to the Nobel Prize".

What should be unacceptable is to present no attempt to *search* for an explanation. Indeed, a paper that just dumps upon the innocent readers a bunch of graphs is not going to make a respectable publication. However, the paper that states that "we have considered this, this and this possible interpretations, and none of them fully explains the data" can only be praised. Similarly, a paper that cheerfully concludes that "our results are consistent with topological-nematic-Majorana-Weyl-whatever" has no right to exist *unless* this conclusion is preceded with an analysis of all *alternative* explanations that are not ground-breaking, along with convincing arguments that these can be safely eliminated.

If you are reading this essay, you probably agree with me. I know many editors do. I know my fellow researchers do. So, if we are all on the same side, who is our evil twin?

Well, we all are. If you submit papers to a high-profile journal, you are probably also guilty (as I am) of having desperately sought a plausible — and ideally nontrivial — explanation of the exper-imental data at hand. Most of us have done this even if we knew in our guts that we were peddling something that cannot be excluded, but is most likely not there.

If you are an editor, you are probably also guilty (as I am) of looking at a submission and thinking: "I, personally, find this an interesting read; but the referees will most likely not find it sufficiently exciting, so let's save time and send it back right now".

If you are a referee, you are probably also guilty (as I am) of thinking that "hey, their pretentious interpretation probably does not make any sense, but who cares – the data are very interesting, so let us write it off as a sales pitch" or, even worse "Hmmm… the experiment looks quite intriguing but they have no idea about the meaning of their observations… therefore, this is not *Nature/Science* material".

Maybe it's time to stop deflecting the blame (the editors won't like it… the reviewers won't like it… and, our favorite: the program managers won't like it), and take responsibility ourselves. If you do not like the inverse Occam's razor cutting into our research, put your money where your mouth is.

There are no evil twins – unless you count human nature (as pertaining to nature of our psyche, not to *Nature*) and its tendency to be excited about something understandably interesting rather than something incomprehensible (and possibly trivial).

When in the brilliant finale of the television series *Quantum Leap* the bartender (who is implied to be the Supreme Being) tells the protagonist, "Sometimes, *'that's the way it is'*, is the best explanation", our soul rebels. We *want* a better explanation! Indeed, '*that's the way it is'* is not an explanation. But, absence of an explanation here and now is more enticing than vigorous hand-waving in the wake of another fashionable scientific trend. This is especially true when this hand waving is clearly aiming at just generating a buzz-word cloud. As Scott DelConte, a New York judge, once observed, ordering a review of an election, "It is more important to do this right than right now".

Let me end this supplication with another quote, misattributed, this time, to Niels Bohr*: there is an infinite number of incorrect theories correctly explaining the finite number of experimental data*. This is precisely why any scientific effort must respect Occam's parsimony.

---

[1] https://www.britannica.com/topic/Occams-razor

[2] Kapitza, P. Viscosity of Liquid Helium below the λ-Point. Nature 141, 74 (1938). https://doi.org/10.1038/141074a0

[3] Of course, in those days Letters to Nature were exactly that -- letters sent in, picked by the editor for its potential interest and published as is without further scrutiny or peer review. Nobody these days would condone such a practice". I agree – nor did it really work that well even then. Of 768 Letters published in Nature in 1938 only 5% were cited more than 30 times, including Ref. 2, and another Nobel paper by F. London.